\documentclass[aps,superscriptaddress,showpacs,nofootinbib,twocolumn]{revtex4}

\usepackage[dvips]{color}                 %%% for colours in the graphs
\usepackage{eucal}

\usepackage{epsfig}

\begin{document}

\title{Asymmetrically Doped Polyacetylene}

\author{Heron Caldas} \email{hcaldas@ufsj.edu.br} \affiliation{Departamento de
  Ci\^{e}ncias Naturais, Universidade Federal de S\~{a}o Jo\~{a}o del Rei,\\
  36301-160, S\~{a}o Jo\~{a}o del Rei, MG, Brazil}

\begin{abstract}
Doped one-dimensional (1D) conjugated polymers, such as polyacetylene, have a conductivity of some metals, like copper. We show that when this polymer is asymmetrically doped, unexpected properties are revealed, when compared to the behavior of the symmetrically standard doped systems (SDS). Depending on the level of imbalance between the chemical potentials of the two involved fermionic species, the polymer can be converted into a 1D partially or fully spin polarized organic conductor.

\end{abstract}


\maketitle

\section{Introduction}

The interest in polyacetylene was revived recently by many reasons. From the theoretical point of view, it can been considered a 1D version of graphene, at least with respect to the fractionalization of the electric charge \cite{Jackiw,Chamon,Franz}. From the experimental and technological sides, the advent of the {\it semiconductor nanotechnology}~\cite{Duan} with new possibilities of developing semiconductor devices made from polyacetylene~\cite{Nature1} strongly motivated the study of transport properties of conducting 1D polymers \cite{Aleshin}. Polyacetylene is a 1D chain of $\rm CH$ groups with alternating single and double bonds. The system is half filled, i.e., has one electron per site and, in the tight binding approximation, it would be a metal. However, due to the interaction of the electrons with the lattice or, the spontaneous Peierls dimerization, the polyacetylene is an insulator. After doping, the conductivity of {\it trans}-polyacetylene is greatly increased, reaching metal-like properties~\cite{Review}. 

The model which describes the electron-phonon interactions is the celebrated Su-Schrieffer-Heeger (SSH) Hamiltonian~\cite{SSH}, whose continuum version is known as the Takayama--Lin-Liu--Maki (TLM) model \cite{TLM}. The TLM model is a relativistic field theory with two-flavor Dirac fermions. We shall use the Gross-Neveu (GN) model \cite{GN} in 1+1 dimensions since its Lagrangian is equivalent to that of the TLM model \cite{CB}, and will provide us with analytic solutions. In the field theory context, the discrete chiral symmetry is spontaneously broken in the GN model (at zero temperature and chemical potential), and there is the dynamical generation of a mass (gap) for the electrons. The (global) external chemical potential $\mu$ is introduced in the theory to represent the extra electrons that are supplied in the system by the doping process. At a critical chemical potential $\mu_c=\frac{\Delta_0}{\sqrt{2}}$, where $\Delta_0$ ($\approx 0.7 {\rm eV}$ for trans-$(\rm CH)_x$) is the constant band-gap, the GN model undergoes a first order phase transition to a symmetry restored (zero gap) phase~\cite{Wolff}. The critical doping concentration~\cite{y} $y_c$ can be related with the critical chemical potential as $y_c=\frac{N}{\pi \hbar v_F} a \mu_c$~\cite{CM}, where $N~(=2)$ is the number of spin degrees of freedom of the (delocalized) $\pi$ electron, $\hbar$ is the Planck's constant divided by $2\pi$, $v_F=k_F\hbar/m$ is the Fermi velocity, $k_F$ is the Fermi wavenumber, $m \equiv \hbar^2/2t_o a^2$, and $a$ ($\cong 1.22 {\rm \AA} $) is the lattice (equilibrium) spacing between the $x$ coordinates of successive ${\rm CH}$ radicals in the undimerized structure. $t_0 (\approx 3~ {\rm eV})$~\cite{Review} is the intercarbon transfer integral for $\pi$ electrons, or simply the ``hopping parameter", resulting in $v_F \approx 10^6 m/s$ for trans-$(\rm CH)_x$, which is of the order of the velocity of the Dirac fermions in graphene~\cite{Graphene}. Employing the GN model in the large-$N$ (mean-field) approximation, previous work~\cite{CM} found a very good agreement with the experimentally found $y_c$ ($\cong 6 \%)$. Recent calculations including finite corrections to $N$ improved the theoretical prediction for $y_c$, and showed that $y_c$ is solely weakly affected by thermal effects~\cite{PRB}.

The novelty is brought about by an ``asymmetrical doping'', defined as an imbalance between the chemical potentials of the electrons with the two possible spin orientations (``up" $\equiv \uparrow$, and ``down" $\equiv \downarrow$) inserted in the system by the doping process. Since the densities of the $\uparrow$ and $\downarrow$ electrons are directly proportional to their chemical potentials, we investigate the consequences of the asymmetrical doping in the magnetic properties of the {\it trans}-polyacetylene.

Experimentally, the chemical potential asymmetry can be achieved by the actuation of an external magnetic field on the system, which breaks the spin-$1/2$ SU(2) symmetry. We show that with low imbalanced doping (with respect to the critical doping), the energy gap is zero, although there is an imbalance between the $\uparrow$ and $\downarrow$ electron populations. The situation changes substantially with strong imbalanced doping, favoring one of the movers from the two spin orientations ($\uparrow$ or $\downarrow$). In this case, the system acquires a small but stable non-zero gap $\Delta_0(\delta \mu) < \Delta_0$, and there is an absence of the not favored spin oriented electron. We argue that this may have observable consequences in the magnetic properties of the asymmetrically doped polyacetylene (ADP).

\section{Model Lagrangian}

The Lagrangian density of the TLM model in the adiabatic approximation (i.e., neglecting the lattice vibrations) is given by

\begin{eqnarray}
{\cal L}_{\rm TLM} &=&
 \sum_{j=1}^N {\psi^j}^\dagger
\left( i  \partial_t - i \hbar v_F \gamma_5 \partial_x - \gamma_0
\Delta(x) 
\right) \psi^j \\
\nonumber
&-&\frac{1}{2 \pi \hbar v_F \lambda_{\rm TLM}} 
\Delta^2(x)\;,
\label{LagTLM}
\end{eqnarray}
where $\psi$ is a two component Dirac spinor $\psi^j= \left(\begin{array}{cc} \psi^j_{L}\\ \psi^j_{R} \end{array} \right)$, representing the ``left-moving" and ``right-moving" electrons close to their Fermi energy, respectively, and $j$ is an internal symmetry index (spin) that determines the effective degeneracy of the fermions. We define $1=\uparrow$, and $2=\downarrow$. The gamma matrices are given in terms of the Pauli matrices, as $\gamma_0=\sigma_1$, and $\gamma_5=-\sigma_3$. $\Delta(x)$ is a (real) gap related to lattice vibrations, $\lambda_{\rm TLM} = \frac{ 2 \alpha^2}{\pi t_0 K}$ is a dimensionless coupling, where $\alpha$ is the $\pi$-electron-phonon coupling constant of the original SSH Hamiltonian, and $K$ is the elastic chain deformation constant.

The equivalence between the TLM and the Gross-Neveu (GN) model, is established by setting $\lambda_{\rm TLM} = \frac{\lambda_{\rm GN}}{N \pi} $. The partition function of the GN model in the imaginary time formalism~\cite{Kapusta} is given by

\begin{equation}
Z=\int [D \psi^{\dagger}][D \psi] exp {\int_0^{\beta} d\tau \int dx~({L}_{\rm GN})},
\label{partfun}
\end{equation}
where $\beta=1/k_BT$, $k_B$ is the Boltzmann constant, and $L_{\rm GN}$ is the Euclidean GN Lagrangian. The ${\rm(CH)_x}$ can be doped after synthesis both by using chemical or electrochemical techniques. If an electrochemical process is being used to add electrons through external circuits, the polymer is being reduced. When the doping removes electrons it is said that the polymer is being oxidized. In the standard way, the chemical potential is introduced in the theory by adding to ${L}_{\rm GN}$ the term $\mu \psi^{\dagger} \psi$, which means that the same quantity of $\uparrow$ and $\downarrow$ electrons is being inserted in the system. The asymmetrical doping is established by adding to ${L}_{\rm GN}$ a term $\mu_{\uparrow} {\psi^\uparrow}^{\dagger}  \psi^{\uparrow} + \mu_{\downarrow} {\psi^\downarrow}^{\dagger}  \psi^{\downarrow}$, with $\mu_\uparrow= \bar \mu + \delta \mu$, and $\mu_\downarrow= \bar \mu - \delta \mu$. We will also choose $\bar \mu=\mu_c$ since, as we will see next, this is the chemical potential at which the minimum of $V_{eff}(\Delta,\mu_{\uparrow,\downarrow})$ jumps (when $\delta \mu=0$) from $\Delta_0$ to $0$~\cite{Wolff}. This allows us to write $L_{\rm GN} \equiv {L}_\uparrow + {L}_\downarrow + L_0$, where ${L}_{\uparrow,\downarrow} =\bar \psi_{\uparrow,\downarrow} \left( -  \gamma_0 \partial_\tau + i \hbar v_F \gamma_1 \partial_x -  \Delta  +\gamma_0 \mu_{\uparrow,\downarrow} \right) \psi_{\uparrow,\downarrow}$, $L_0=-\frac{N}{ 2 \hbar v_F \lambda_{\rm GN}} \Delta^2$, $\bar \psi \equiv \psi^{\dagger} \gamma_0 = \psi^{\dagger} \sigma_1$, and $\gamma_1=i \sigma_2$. Integrating over the fermion fields we obtain

\begin{eqnarray}
Z=e^{\beta V \left( -\frac{1}{ \hbar v_F \lambda_{\rm GN}} \Delta^2 \right)}~\Pi_{j=1}^2 det D_j,
\nonumber
\label{partfun2}
\end{eqnarray}
where $D_{1,2}=-i \beta [(-i \omega_n + \mu_{\uparrow,\downarrow}) -v_F\gamma_0 \gamma_1 p-\gamma_0 \Delta]$, and $\omega_n=(2n+1)\pi T$ are the Matsubara frequencies for fermions. Summing over $n$ and carrying out the determinantal operation, we find

\begin{eqnarray}
\ln Z = -\beta V \left( \frac{1}{ \hbar v_F \lambda_{\rm GN}} \Delta^2 \right) + \frac{V}{2}  \int^{+\infty}_{-\infty}{\frac{dp}{2\pi \hbar}}~ \Big[ \\
\nonumber 
+ \beta E_{\uparrow}^{+} + 2 \ln \left(1+e^{-\beta E_\uparrow^+}\right)+ \beta E_{\uparrow}^{-} + 2 \ln \left(1+e^{-\beta E_\uparrow^-}\right)\\
\nonumber
+ \beta E_{\downarrow}^{+} + 2 \ln \left(1+e^{-\beta E_\downarrow^{+}}\right)+ \beta E_{\downarrow}^{-} + 2 \ln \left(1+e^{-\beta E_\downarrow^-}\right) \Big],
\label{partfun3}
\end{eqnarray}
where $V$ is the 1D ``volume", $E_{\uparrow,\downarrow}^{\pm} \equiv E_p \pm \mu_{\uparrow,\downarrow}$, and $E_p=\sqrt{v_F^2 p^2+\Delta^2}$. The ``effective potential" for a constant $\Delta$ field is given by $V_{eff}=-\frac{k_BT}{V} \ln Z$, then 

\begin{eqnarray}
V_{eff}&=&\frac{1}{ \hbar v_F \lambda_{\rm GN}} \Delta^2 -  \int^{+\infty}_{-\infty}{\frac{dp}{2\pi \hbar}}~ \Big[ 2 E_p\\
\nonumber 
&+& k_B T \ln \left(1+e^{-\beta E_\uparrow^+}\right)+ k_B T \ln \left(1+e^{-\beta E_\uparrow^-}\right)\\
\nonumber
&+&  k_B T \ln \left(1+e^{-\beta E_\downarrow^+}\right) + k_B T \ln \left(1+e^{-\beta E_\downarrow^-}\right) \Big].
\label{poteff}
\end{eqnarray}

The zero temperature and chemical potentials $V_{eff}$ is written as

\begin{eqnarray}
V_{eff}(\Delta)=\frac{1}{ \hbar v_F \lambda_{\rm GN}} \Delta^2 -  \int^{\Lambda}_0{\frac{dp}{\pi \hbar}}~ 2 E_p,
\label{poteff2}
\end{eqnarray}
where we have introduced a momentum cutoff $\Lambda$ to regulate the vacuum (divergent) part of $V_{eff}(\Delta)$. It is easy to see that $\delta \mu$ has lifted the degeneracy of the conduction $(+)$ and valence $(-)$ bands in the matter part of $V_{eff}$. The renormalized effective potential turns out to be

\begin{eqnarray}
V_{eff}(\Delta)=\frac{\Delta^2}{ \hbar v_F } \left(\frac{1}{\lambda_{\rm GN}} - \frac{3}{2 \pi} \right) + \frac{\Delta^2}{ \hbar v_F \pi} \ln \left( \frac{\Delta}{m_F} \right),
\label{poteff3}
\end{eqnarray}
where $m_F$ is an arbitrary renormalization scale. The minimization of $V_{eff}(\Delta)$ with respect to $\Delta$ gives the well-known result~\cite{GN}

\begin{equation}
\Delta_0 = m_F e^{ 1- \frac {\pi}{\lambda_{\rm GN} }} .
\label{deltaopt}
\end{equation}

At finite chemical potentials and in the zero temperature limit, we have

\begin{eqnarray}
V_{eff}(\Delta,\mu_{\uparrow,\downarrow})=\frac{1}{ \hbar v_F \lambda_{\rm GN}} \Delta^2 -  \int^{\Lambda}_0{\frac{dp}{\pi \hbar}}~ 2 E_p\\
\nonumber 
+\int_0^{p_{F}^\uparrow} \frac{dp}{\pi \hbar} (E_p - \mu_\uparrow) + \int_0^{p_{F}^\downarrow} \frac{dp}{\pi \hbar} (E_p - \mu_\downarrow),
\label{poteff4}
\end{eqnarray}
where $p_{F}^{\uparrow,\downarrow}=\frac{1}{v_{F}}\sqrt{\mu_{\uparrow,\downarrow}^2-\Delta^2}$ is the Fermi momentum of the $\uparrow$($\downarrow$) moving electron. We integrate in $p$, observing that the renormalization is the same as before, to obtain

\begin{eqnarray}
&V_{eff}&(\Delta,\mu_{\uparrow,\downarrow})=\frac{\Delta^2}{ \hbar v_F } \left(\frac{1}{\lambda_{\rm GN}} - \frac{3}{2 \pi} \right) + \frac{\Delta^2}{ \pi \hbar v_F } \ln \left( \frac{\Delta}{m_F} \right)\\
\nonumber
&+& \frac{\Theta_1}{2\pi \hbar v_F }\left[\Delta^2 \ln \left(\frac{\mu_\uparrow + \sqrt{\mu_\uparrow^2-\Delta^2}}{\Delta} \right) -\mu_\uparrow \sqrt{\mu_\uparrow^2-\Delta^2} \right]\\
\nonumber
&+& \frac{\Theta_2}{2\pi \hbar v_F }\left[\Delta^2 \ln \left(\frac{\mu_\downarrow + \sqrt{\mu_\downarrow^2-\Delta^2}}{\Delta} \right) -\mu_\downarrow \sqrt{\mu_\downarrow^2-\Delta^2} \right],
\label{potefff}
\end{eqnarray}
where $\Theta_{1,2}=\Theta(\mu_{\uparrow,\downarrow}^2-\Delta^2)$ is the step function, defined as $\Theta(x)=0$, for $x<0$, and $\Theta(x)=1$, for $x>0$. Minimizing $V_{eff}(\Delta,\mu_{\uparrow,\downarrow})$ with respect to $\Delta$ yields the trivial solution ($\Delta=0$), and

\begin{eqnarray}
\label{min2}
\ln \left( \frac{\Delta}{\Delta_0} \right) &+&  \frac{\Theta_1}{2} \ln \left(\frac{\mu_\uparrow + \sqrt{\mu_\uparrow^2-\Delta^2}}{\Delta} \right) \\
\nonumber
&+& \frac{\Theta_2}{2} \ln \left( \frac{\mu_\downarrow + \sqrt{\mu_\downarrow^2-\Delta^2}}{\Delta} \right)=0,
\end{eqnarray}
where we have made use of Eq.~(\ref{deltaopt}) to eliminate $m_F$ in the gap equation. The ground state is determined by jointly finding the solutions of the equation above $\Delta_{0}(\delta \mu)$, and the analysis of the effective potential at the minimum, $V_{eff}(\Delta_{0}(\delta \mu))$. In the SDS, defined for $\delta \mu=0$, the solution $\Delta=\Delta_0$ represents a minimum of $V_{eff}(\Delta,\mu)$ while $\mu < \mu_c$, (where the critical chemical potential, $\mu_c$, is obtained by $V_{eff}(\Delta=0,\mu_{c})=V_{eff}(\Delta=\Delta_0,\mu_c)$), and $\Delta=0$ is a local maximum. When $\mu \geq \mu_c$, $\Delta(\mu \geq \mu_c)$ is a local maximum and $\Delta=0$ is turned into a minimum through a first order phase transition~\cite{Wolff,CM}, agreeing with experiment~\cite{Fernando}. Thus, the gap as a function of the chemical potential in the SDS has the following expression:

\begin{equation}
\Delta_0(\mu)=\Theta(\mu_c-\mu) \Delta_0.
\label{gapmu}
\end{equation}

\begin{figure}[htb]
  \vspace{0.5cm}
  \epsfig{figure=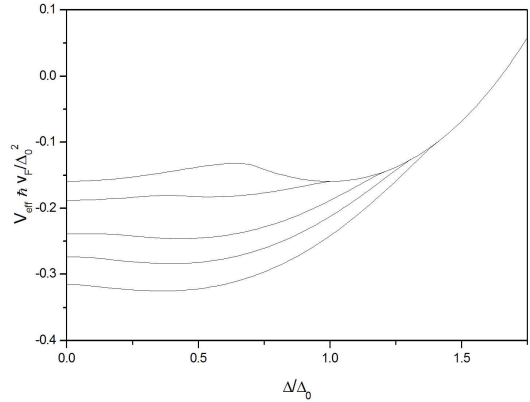,angle=0,width=8cm}
\caption[]{\label{Veff} The top curve is the effective potential of Eq.~(9) for $\delta \mu/\Delta_0=0$, and $\mu=\mu_c$. The second curve from top to bottom is for $\delta \mu/\Delta_0=0.3$, with the minimum of $V_{eff}$ at $\Delta=0$. The following curves are for $\delta \mu/\Delta_0=0.5$, $\delta \mu/\Delta_0=0.6$ and $\delta \mu/\Delta_0=0.7$. There are stable gaps now at $\Delta_0(\delta \mu=0.5 \Delta_0)>\Delta_0(\delta \mu=0.6 \Delta_0)>\Delta_0(\delta \mu=0.7 \Delta_0)$ all of them agreeing with the solution of Eq.~(\ref{min4}).}
\end{figure}

The situation changes significantly in an asymmetrically doped system, where $\Delta=0$ still represents the minimum of $V_{eff}(\Delta,\mu_{\uparrow,\downarrow})$ for $\delta \mu < \delta \mu_c = 0.38 \Delta_0$, where $\delta \mu_c$ is obtained from the equality $V_{eff}(\Delta=0,\mu_{\uparrow,\downarrow}(\delta \mu_c))=V_{eff}(\Delta=\Delta_0(\delta \mu_c),\mu_{\uparrow,\downarrow}(\delta \mu_c))$, with $\Delta_0(\delta \mu_c)$ being the solution of Eq.~(\ref{min2}) for $\mu_{\uparrow,\downarrow}(\delta \mu_c)$. For $\delta \mu/\Delta_0 \geq \delta \mu_c$, the $\Theta_{2}$ function prevents the ``$\downarrow$" term in the effective potential of participating in the minimum. Thus, Eq.~(\ref{min2}) can be rewritten as

\begin{equation}
\label{min4}
\Delta^4-2\mu_\uparrow \Delta_0^2 \Delta + \Delta_0^4=0.
\end{equation}
We show in Fig.~(\ref{Veff}) the (non-dimensional) function $V_{eff}\hbar v_F/\Delta_0^2$ for $\delta \mu/\Delta_0 =0$, $\delta \mu/\Delta_0 = 0.3$, $\delta \mu/\Delta_0 = 0.5$, $\delta \mu/\Delta_0 = 0.6$, and $\delta \mu/\Delta_0 = 0.7$, as a function of $\Delta/\Delta_0$. We see that the minimum of $V_{eff}\hbar v_F/\Delta_0^2$ for $\delta \mu < \delta \mu_c$ is at $\Delta=0$, while for $\delta \mu > \delta \mu_c$, a narrow gap $\Delta(\delta \mu)$ ``appears" through a first order (quantum) phase transition. However, this transition now is from $\Delta=0$ to $\Delta(\delta \mu)\neq0$, i.e., is in opposite direction from that observed in SDS ($\delta \mu=0$), as a function of $\mu$. The behavior of $V_{eff}\hbar v_F/\Delta_0^2$ shows that the gap will eventually vanish for very high asymmetry, or $\delta \mu >> \delta \mu_c$.

\section{Magnetic Properties}

As we mentioned, the imbalance in the chemical potentials of the $\uparrow$ and $\downarrow$ electrons can be caused by the application of a static magnetic field $\vec{B}=B_0 \vec{k}$, perpendicular to the wire, with a Zeeman energy $\Delta E= S_z g \mu_B B_0$~\cite{Madelung,Kittel}, where $S_z=\pm 1/2$, $g \approx 2$ is the effective $g-$factor, $\mu_B=e \hbar/2m \approx  5.788 \times 10^{-5} ~{\rm eV}~ {\rm T}^{-1}$ is the Bohr magneton, $m$ is the bare electron mass, and $B_0$ is the magnetic field strength. In the present case $\delta \mu =|\Delta E|=\mu_B B_0$. 

The number densities $n_{\uparrow,\downarrow}= -\frac{\partial}{\partial \mu_{\uparrow,\downarrow}} V_{eff}(\Delta,\mu_{\uparrow,\downarrow})$ are obviously imbalanced due to the asymmetry between $\mu_\uparrow$ and $\mu_\downarrow$, and will depend on $\delta \mu$. Before the critical asymmetric doping, $\Delta_0(\delta \mu)=0$, and the densities read

\begin{equation}
n_{\uparrow,\downarrow}(\delta \mu < \delta \mu_c)= \frac{1}{\pi \hbar v_F } \mu_{\uparrow,\downarrow}.
\label{nd}
\end{equation}
For such a low imbalance, compared to the critical chemical potential asymmetry $\delta \mu_c$, the total number density, $n_T$, is the same as in the symmetric limit or, in other words, is independent of the applied field:

\begin{equation}
n_T(\delta \mu < \delta \mu_c)=n_{\uparrow}+n_{\downarrow}=2n=\frac{2}{\pi \hbar v_F } \mu_c.
\label{nT}
\end{equation}
This partial spin polarization results in a net magnetization~\cite{Kittel} of the chain

\begin{eqnarray}
M(\delta \mu < \delta \mu_c)&=& \mu_B (n_\uparrow - n_\downarrow)= \frac{2 \mu_B}{\pi \hbar v_F } \delta \mu\\
\nonumber
&=&\frac{2 \mu_B^2}{\pi \hbar v_F } B_0 \equiv M_0.
\label{nm}
\end{eqnarray}
The magnetic susceptibility in the low asymmetrical doping regime is

\begin{equation}
\chi(\delta \mu < \delta \mu_c)=\frac{\partial M}{\partial B_0}=\frac{2 \mu_B^2}{\pi \hbar v_F },
\label{ms}
\end{equation}
which is exactly the Pauli expression for noninteracting electrons.

Increasing the asymmetry beyond the critical value means that, effectively, only $\uparrow$-moving electrons of the conduction band are present in the system. In terms of this ``higher" chemical potential asymmetry the densities are

\begin{equation}
n_{\uparrow}(\delta \mu > \delta \mu_c)= \frac{1}{\pi \hbar v_F }\sqrt{\mu_{\uparrow}^2-\Delta^2},
\label{nd3}
\end{equation}
where $\Delta=\Delta(B_0)$ in the equation above is the solution of Eq.~(\ref{min4}) for a given $\mu_{\uparrow}$, and   

\begin{equation}
n_{\downarrow}(\delta \mu > \delta \mu_c)= 0,
\label{nd4}
\end{equation}
meaning that the system is fully polarized. The total number density is now

\begin{equation}
n_T(\delta \mu > \delta \mu_c)=n_{\uparrow}=\frac{1}{\pi \hbar v_F }\sqrt{(\mu_c+ \mu_B B_0)^2-\Delta^2}.
\label{nT2}
\end{equation}
In the fully spin polarization regime the magnetization of the chain reads

\begin{eqnarray}
M(\delta \mu > \delta \mu_c)&=& \mu_B n_\uparrow \\
\nonumber
&=&\frac{\mu_B}{\pi \hbar v_F }\sqrt{(\mu_c+ \mu_B B_0)^2-\Delta^2} > M_0.
\label{nm2}
\end{eqnarray}
The magnetic susceptibility in this regime is given by a rather involved expression

\begin{equation}
\chi(\delta \mu > \delta \mu_c)=\frac{\mu_B}{\pi \hbar v_F } \frac{(\mu_c+\mu_B B_0)\mu_B-\Delta\frac{\partial\Delta}{\partial B_0}}{\sqrt{(\mu_c+ \mu_B B_0)^2-\Delta^2}},
\label{ms2}
\end{equation}
that clearly does not have the simple form of the Pauli susceptibility.

Finally we find the critical magnetic field separating the partially and fully polarized magnetic phases $\delta \mu_c = 0.38 \Delta_0 = \mu_B B_{0,c}$, yielding

\begin{equation}
B_{0,c} \approx 4.6 ~{\rm k T},
\label{cf}
\end{equation}
which is a magnetic field of very high intensity, compared to the maximum current laboratory values, that are up to $ 20~{\rm T}$ (DC-magnetic fields), and up to $331~{\rm T}$ (mega-gauss fields)~\cite{lab}.
\\

\section{Summary}

To summarize, we have investigated the mean-field zero-temperature phase diagram of a 1D polymer upon asymmetric doping. Our main finding is that an ADP can be converted into a partially or fully magnetized organic wire, depending on the level of doping imbalance, i.e., the difference between the $\uparrow$ and $\downarrow$ electron densities, induced by an external magnetic field. Although the (critical) magnetic field $B_{0,c}$ necessary for a fully magnetization of the polyacetylene be of very high magnitude, we have seen that any non-zero $B_{0}$ produces a net magnetization. This doable partial spin polarization can be a manifestation of the interesting ``itinerant magnetism'', for which conduction electrons are responsible. It would also be desirable to study the polarization mechanism in the trans-${(\rm CH)_x}$ at finite temperature, and the subsequent calculation of thermal effects on the magnetization, critical magnetic field, and susceptibility, within the field theory approach. We hope to address these issues in future.

\section{Acknowledgments}

The author acknowledges partial support by CNPq-Brazil and FAPEMIG. I am grateful to Dr. A. L. Mota for stimulating conversations. I also would like to thank the Institute for Nuclear Theory, University of Washington, and the Quarks, Hadron and Nuclei group, Physics Department, University of Maryland, at which parts of this work were made, for their hospitality. Finally I would like to express my appreciation to the Referee whose careful reading of and valuable remarks on the manuscript undoubtedly led to a better paper.


\begin{thebibliography}{99}
\bibitem{y}The doping concentration $y$ is a dimensionless parameter, defined as the number of dopants per carbon atom~\cite{CM}.
\end{thebibliography}
\end{document}